\documentclass[12pt]{article} 

\newcommand{\ket}[1]{\left| #1 \right>}  
\newcommand{\bra}[1]{\left< #1 \right|} 
\newcommand{\brahket}[3]{\left< #1 \left| #2 \right| #3 \right>}

\title{\bf Recoherence in the entanglement dynamics and classical orbits 
in the $N$-atom Jaynes-Cummings model} 
\author{R. M. Angelo$^{(1)}$, K. Furuya$^{(1)}$, M. C. Nemes$^{(2)}$,  
{\small and} G. Q. Pellegrino$^{(3)}$\\ \ \\ 
{$^{(1)}$ 
\normalsize \em Instituto de F\'{\i}sica ``Gleb Wataghin'', Universidade  
Estadual de Campinas}\\ 
{\normalsize \em 13083-970 Campinas -- SP, Brasil}\\ 
{$^{(2)}$ 
\normalsize \em Departamento de F\'{\i}sica -- ICEx, Universidade Federal de  
Minas Gerais}\\ 
{\normalsize \em 31270-901 Belo Horizonte -- MG, Brasil}\\ 
{$^{(3)}$ 
\normalsize \em Departamento de Matem\'atica, Universidade Federal de  
S\~ao Carlos}\\ 
{\normalsize \em 13560-970 S\~ao Carlos -- SP, Brasil}} 
\date{\today} 
\begin{document} 
\maketitle 
\begin{abstract} 
   The rise in linear entropy of a subsystem in the $N$-atom Jaynes-Cummings  
model is shown to be strongly influenced by the {\bf shape} of the classical 
orbits of  the underlying classical phase space: we find a one-to-one  
correspondence between maxima (minima) of the linear entropy and maxima  
(minima) of the expectation value of atomic excitation $J_z$. Since the  
expectation value of this operator can be viewed as related to the orbit radius in the 
classical phase space projection associated to the atomic degree of freedom, the
proximity of the quantum wave packet to this atomic  phase space borderline produces a
maximum rate of entanglement. The   consequence of this fact for initial conditions
centered at periodic orbits   in regular regions is a clear periodic recoherence. For
chaotic situations the same phenomenon (proximity of the atomic phase space
borderline) is in general responsible for oscillations in the entanglement
properties. 
\end{abstract} 
\newpage 
\setcounter{page}{2} 
\section{INTRODUCTION} 
   The importance of studying in detail the decoherence process is twofold.  
Firstly it may be viewed as a key to the understanding of some of the striking  
differences between the quantum and classical description of the world such as ``the 
non existence at the classical level of the majority of states allowed  by quantum 
mechanics'' \cite{Einstein}. The decoherence process is believed  to be the agent 
which eliminates interference between two or more  macroscopically separated 
localized states \cite{decoherence}. Secondly, given the impressive technological 
advances in several experimental areas (Quantum Optics, Condensed Matter and Atomic 
Physics, etc.), it is nowadays  possible to realize a system of two interacting 
degrees of freedom and watch the time evolution of the corresponding entanglement 
process \cite{Brune}.  It is therefore also of importance to understand  the 
entanglement process  in simple Hamiltonian systems. Two degrees of freedom 
Hamiltonian systems often  present a very rich dynamics which in many cases is not 
yet completely  understood from a general point of view. In particular, if the 
interaction  is nonlinear the system may present chaotic behavior in the classical 
limit.  The consequences of this fact to the quantum dynamics is yet an unsettled  
issue. A step in this direction was taken a few years ago, as it was  
conjectured that ``the rate of entropy production can be used as an  
intrinsically quantum test of the chaotic versus regular nature of the  
evolution'' \cite{zurek95}. The idea has been tested in some models  
\cite{Shiokawa,nos}. More specifically, in the  
context of the $N$-atom Jaynes-Cummings model, the reduced density linear 
entropy (or idempotency defect) has been used as a measure of the  
entanglement of the quantum subsystems. For a given classical energy, initial 
conditions for the quantum states are prepared as coherent wave packets centered at
regular and chaotic regions of the classical phase space. For short times, a fast
increase in decoherence for chaotic initial conditions is found when compared to
regular ones. Typically the linear entropy in this model rises from zero to a
plateau.  
 
 In the present contribution we show that this rise 
is strongly influenced by the {\bf shape} of the atomic projection of classical 
orbits. There is a clear correlation between the rate of increase of linear entropy
and the increase of the expectation value of the atomic  excitation $\langle J_z
\rangle(t)$ for short enough times. In classical terms, this can be  visualized as
follows: $\langle J_z \rangle(t)$  is a measure of the instantaneous radius of the
projection of the   trajectory in the atomic phase space.  This atomic phase space is
limited, the maximum radius corresponds to $\sqrt{4J}=\sqrt{2N}$, meaning therefore
that the maximum rate of entanglement can be directly associated to the proximity of
this borderline. If the quantum evolution is such that the initial wave packet is
centered at a classically regular region of the atomic phase space, in particular on a
(non  circular!) periodic orbit, the phenomenon is most conspicuous: as a function of
time, the wave packet exhibits partial recoherences, well marked oscillations in the
linear entropy superimposed to a steady increase, which can be immediately associated
to smallest and largest distances from the classical border (largest or smallest
$\langle J_z \rangle(t)$). If, on the other hand, the initial wave packet is centered
on a classically chaotic region of the phase space, the recoherence phenomenon is of
course not necessarily periodic in general. However we still find a connection between
maxima and minima of the rate of linear entropy and $\langle J_z \rangle(t)$. As to be
expected, this simple relation becomes less clear as quantum effects set in; in
particular, for the chaotic region this happens much faster than for the regular one
in a mixed phase space. Such effects are most marked at sufficiently long times, when
a plateau in the linear entropy is reached. We show that at these times the spin
projection of the wave packet becomes totally delocalized in phase space. This takes
place for initial conditions centered at regular {\bf and} chaotic regions, the
regular cases taking longer times to achieve this  delocalization. 
  
   In section II we present the model and the technical tools to be used 
in the following sections to analyze the entanglement property, connecting it to the 
classical regularity and chaos in a mixed regimen. Section III is 
devoted to the presentation of analysis and results in regular regions, whereas 
section IV deals with chaotic regions. A summary is given in section V. 
 
\section{THE MODEL} 
 
  We consider the $N$-atom Jaynes-Cummings model \cite{NJCM} whose  
Hamiltonian is given by 
\begin{equation} 
H=\hbar\omega_0 a^+a+\varepsilon J_z+\frac{G}{\sqrt{2J}}\left( {aJ_++a^+J_-}  
\right) + \frac{G'}{\sqrt{2J}}\left( {a^+J_++aJ_-} \right) \label{Hq} 
\end{equation} 
where the first term corresponds to the energy of the free single-mode  
quantized field with frequency $\omega_0$, described by the creation  
(annihilation) operators $a^+$ ($a$); the second term corresponds to the  
energy of the $N=2J$ two-level atoms with energy separation $\hbar\varepsilon$, and 
the operators $J_z$, $J_{\pm}= J_x \pm iJ_y$ are the usual angular momentum  
ones corresponding to the group $SU(2)$. These operators satisfy the well known 
commutation  relations: 
\begin{eqnarray} 
\left[ a, a^{+} \right] & = & 1, \nonumber \\ 
\left[ J_i,J_j \right]  & = & i\hbar J_k 
\nonumber 
\end{eqnarray}  
with the indices $(i,j,k)$ forming any cyclic permutation of $(x,y,z)$. The last 
two terms in eq. (\ref{Hq}) represent the interaction energy between the atomic 
system and the single-mode field.

The above Hamiltonian is a generalized version of the usual
$N$-atom Jaynes-Cummings model to which we have added a Counter-Rotating 
Wave (CRW) term, with coupling constant $G^{'}$. The so-called Rotating Wave 
Approximation (RWA) has been much in use for the proposals of mesoscopic 
superpositions of collective atomic states \cite{ATOMCAT,Recamier00} and in 
the generation of multiparticle entangled states in cavity QED experiments 
\cite{Rauschenbeutel00}.

The $N$-atom JCM can also be realized in systems where trapped ions interact 
with laser fields. In that situation the bosonic operators $a$ and $a^+$ 
describe the vibrational motion of the ions in the trap. Recently, reversible 
entanglement of ions has been proposed \cite{Sorensen00}, and experimental 
realizations have been reported in the cases of two and four ions 
\cite{Turchette98}, with possible extensions to several ions \cite{Roos99}. 
In the trapped-ion system there is the possibility of generating both RW and 
CRW types of interaction by means of either a bichromatic laser excitation 
--- with both $G$ and $G^{'}$ different from zero --- or appropriate excitation 
modes \cite{Cirac96}. In the latter case, there would be the possibility to test 
the effects of RW ($G^{'} = 0$) and CRW ($G = 0$) terms separately or together.

In our model we also add the supposition that the two coupling constants can be 
independently varied. When one of the coupling constants $G$ or $G^{'}$ is set to 
zero, we have an integrable system. For $G^{'}=0$ ($G=0$) there is an additional 
conserved quantity, namely the total excitation $P=J_z/\hbar+a^+a$ 
(the relative excitation $P^{'}=J_z/\hbar-a^+a$). The system is otherwise  
nonintegrable and known to exhibit chaotic behavior in the classical limit 
\cite{Milonni98}. In both cases, the quantum dynamics will produce entanglement, 
due to the coupling terms. This essentially quantum property can be quantified, 
e.g. by means of the linear entropy (or idempotency defect) associated to the 
reduced density matrices of the atomic or field subsystems (from this point on 
we shall set $\hbar =1$), 
\begin{eqnarray} 
\delta(t)=1-Tr_i\left(\rho_i(t)^2\right),    \label{ID} 
\end{eqnarray} 
where 
\begin{eqnarray} 
\rho_i(t) = Tr_j\left(\ket{\psi(t)} \bra{\psi(t)}\right)      \label{RD} 
\end{eqnarray} 
($i$ and $j$ stand for the atomic and field subsystems, $i\neq j$). The quantities 
expressed in eqs. (\ref{ID}) and (\ref{RD}) are to be calculated using quantum states
evolving in time under the action of Hamiltonian (\ref{Hq}). To this end, the initial
conditions chosen in the present study are such that  
\begin{equation} 
\ket{\psi(0)} = \ket{w} \otimes \ket{v} \equiv \ket{wv} \label{ic} 
\end{equation} 
where $\ket{w}$ ( $\ket{v}$ ) are atomic (field) coherent states given 
by \cite{klauder85} 
\begin{equation} 
\ket{w} = \left( {1+w\bar w} \right)^{-J}e^{wJ_+} \ket{J, -J} \label{SPIN} 
\end{equation} 
\begin{equation} 
\ket{v} = e^{-v\bar v/2}e^{vb^+} \ket{0} \label{FIELD} 
\end{equation} 
with  
\begin{equation} 
w = {{p_a+iq_a} \over {\sqrt {4J-\left( {p_a^2+q_a^2} \right)}}}, 
\end{equation} 
\begin{equation} 
v = {1 \over {\sqrt 2}}\left( {p_f+iq_f} \right); 
\end{equation} 
$\ket{J,-J}$ being the state with spin $J$ and $J_z = -J$, $\ket{0}$  
being the harmonic oscillator ground  state, and $p_a,q_a,p_f,q_f$ describing the phase
space of the system into consideration. Generation of atomic coherent states in 
$N$-atom systems has been proposed in \cite{ATOMCAT} and analyzed 
in \cite{Recamier00}.
 
  The classical Hamiltonian corresponding to eq. (\ref{Hq}) can be obtained by a 
standard procedure as \cite{aguiar92} 
\begin{equation} 
{\cal H}(v,v^{*},w,w^{*}) \equiv \brahket{wv}{H}{wv}. 
\end{equation} 
${\cal H}(v,v^{*},w,w^{*})$ can be rewritten in terms of the phase space 
variables, reading 
\begin{eqnarray} 
{\cal H}(q_a,p_a,q_f,p_f) & = & {{\omega_0} \over 2}\left( {p_f^2+q_f^2} \right)+ 
{\varepsilon  \over 2}\left( {p_a^2+q_a^2 -2J} \right) + \nonumber \\ 
 & & + {\sqrt{1- {{p_a^2+q_a^2}\over{4J}} }} \left( {G_+p_ap_f+G_-q_aq_f} 
\right), \label{Hcl} 
\end{eqnarray} 
where $G_{\pm} = G \pm G'$. 
 
  The corresponding equations of motion are then given by 
\begin{equation} 
\label{EM} 
\begin{array}{l} 
\displaystyle{ 
\dot{q_a} = - \frac{\partial {\cal H}}{\partial p_a} = 
-\varepsilon p_{a} - G_{+}p_{f} 
\sqrt{1-\frac{q_{a}^{2}+p_{a}^{2}}{4J}}+\frac{p_{a}}{\sqrt{4J}} 
\frac{ \left( G_{+}p_{a}p_{f}+G_{-}q_{a}q_{f} \right) }{\sqrt{4J-(q_{a}^{2}+ 
p_{a}^{2})}}} \\ 
\displaystyle{ 
\dot{q_f} = - \frac{\partial {\cal H}}{\partial p_f} = 
-\omega_{0}p_{f} - G_{+}p_{a} 
\sqrt{1-\frac{q_{a}^{2}+p_{a}^{2}}{4J}}} \\ 
\displaystyle{ 
\dot{p_a} = \frac{\partial {\cal H}}{\partial q_a} = 
\varepsilon q_{a} + G_{-}q_{f} 
\sqrt{1-\frac{q_{a}^{2}+p_{a}^{2}}{4J}}-\frac{q_{a}}{\sqrt{4J}} 
\frac{ \left( G_{+}p_{a}p_{f}+G_{-}q_{a}q_{f} \right) }{\sqrt{4J-(q_{a}^{2}+ 
p_{a}^{2})}}} \\ 
\displaystyle{ 
\dot{p_f} = \frac{\partial {\cal H}}{\partial q_f} = 
\omega_{0}q_{f} + G_{-}q_{a} 
\sqrt{1-\frac{q_{a}^{2}+p_{a}^{2}}{4J}}} \; . 
\end{array}  
\end{equation}  
 
  It is important to note the restriction in energy to be shared in the 
classical counterpart of the atomic degree of freedom, i.e., 
\begin{equation} 
 {p_a}^2+{q_a}^2 \le 4J. \label{Bd} 
\end{equation} 
This represents a fundamental difference between these two degrees of 
freedom. From the quantum point of view, the atomic degree of freedom 
is associated to a finite dimensional Hilbert space whereas the harmonic
oscillator  to a Hilbert space of infinite dimension. 
 
   Except when otherwise noted, in all cases to be presented here the parameter values
we chose are $J=9/2$, $G=0.5$ and $G^{'}=0.2$, and $\varepsilon = \omega_0 = 1$ (as
shown in Ref. \cite{lewenkopf91}, for  these parameter values the nearest neighbor
distribution of the energy level spacings is of a GOE type). For the classical
Hamiltonian in eq. (\ref{Hcl}) we chose an energy value $E=8.5$ for which the
Poincar\'e section ($q_f=0$) is shown in Fig. 1a. Note the presence of two islands,
one of them of considerable  relative size. In the center of such islands are located
the two shortest stable periodic orbits whose atomic projections are shown in Fig. 1b.
The symbols in the Poincar\'e section (Fig. 1a) represent the center of the initial
quantum wave packets whose time evolution we will  show in the next section. For the
sake of comparison we also study the classical energy $E=35$ whose Poincar\'e section
is shown in Fig. 2a. The size of the many regular islands is comparatively much
smaller. Both chosen energies are larger than the limit given by eq. (\ref{Bd}),
therefore the entire atomic phase space is energetically available for all initial
conditions in the Poincar\'e sections. Moreover, the fact that the energies are large
enough together with the limitation imposed by eq. (\ref{Bd}) have dramatic
consequences on the form of the projections of the periodic orbits onto the atomic
phase space. One can see an example of this in Fig. 2b (this particular point has been
explored in Refs. \cite{aguiar92,aguiar91}). 
 
\section{ENTANGLEMENT DYNAMICS IN REGULAR REGIONS}  
 
  The linear entropy $\delta(t)$ corresponding to the reduced density matrix for the
atomic subsystem is shown in Fig. 3a for initial conditions of the type given in eq.
(\ref{ic}), where the center of these wave packets are on the periodic orbits
indicated in Fig. 1a. Note that the linear entropy not only increases but exhibits a
well marked oscillatory behavior, indicating that the system recovers coherence in a
periodic way, for sufficiently short times. Interestingly enough, however, the period
of recoherences observed in the linear entropy do {\bf not} correspond to the period
$\tau$ of the periodic orbit in question. It is {\bf smaller}, roughly $\tau/2$. This
indicates that another property of these orbits may be playing a role. This property
is the {\bf shape} of the spin (atomic) projection of the orbit.  Note the strong
correlation between the maximum growth in linear entropy and the closest approach to
the classical atomic phase space border. This happens in the cases shown here and in
all cases we analyzed, for  short enough times. A natural question arises as to the
representation  dependence of our  explanation. In order to clarify this point we use
a  representation independent quantity which is intimately connected to the radius of
the orbit projection, the expectation value of the operator $J_z$ \cite{note}. In Fig.
3a we also show $\frac{d\delta}{dt}$ and $Tr(J_z\rho_a(t))$. Note that the first four
maxima in $\langle J_z \rangle (t)$ correspond to a very good accuracy to maxima in
$\frac{d\delta}{dt}$, indicating that whenever the ``radius'' is maximum, a maximum
growth in linear entropy is found. It is in this sense that we conclude that the
recoherences, purity gains found in the time evolution of quantum coherent wave
packets initially centered at the regular region, are related to the  proximity of the
classical atomic phase space border. Of course when the initial wave packet is
centered on a periodic orbit located in a smaller island, as is the case of the second
orbit in Fig. 1b, the chaotic vicinities also play an important role and the
established connection between $\frac{d\delta}{dt}$ and $Tr(J_z\rho_a(t))$ is less
pronounced, as shown in Fig. 3b. Due to the shape of the orbit, in this case the
proximity to the center and hence the minima of both functions are more clearly seen
to be connected. We next consider two more cases which independently corroborate these
findings. Consider first the Poincar\'e section in Fig. 2a and the initial condition
for the quantum evolution centered at the circle  marked in the section, and the spin
projection of the periodic orbit shown in Fig. 2b. In this case we note that the
amount of time spent in the vicinities of the phase space border (performing the
``loop'' in Fig. 2b) is large compared to the period of the orbit. Note that during
this time, $\frac{d\delta}{dt}$ is positive and the linear entropy grows (see Fig.
3c). Also, for the subsequent times where the orbit quickly crosses from one ``loop''
to the other, it approaches the center of the atomic phase space and correspondingly
$\frac{d\delta}{dt}$ becomes negative and the linear entropy decreases. Since the time
of growth is much larger than the one of decreasing, the linear entropy reaches the
plateau faster.  
 
On the other hand, in the integrable case $G^{'}=0$, the spin (atomic) projection of 
the periodic orbits are circular. In this case, if the argument is valid, one should
expect no oscillations in $\delta(t)$. This is in fact the case, as shown in Fig. 4
where one sees, moreover, that it takes a longer time for the linear entropy to reach
the plateau, due to the considerable (and constant) distance between the orbit and the
atomic phase space border.   
 
\section{ENTANGLEMENT DYNAMICS IN THE CHAOTIC REGION}  
 
  As discussed in Ref. \cite{nos} the entanglement in the chaotic 
region is in general faster than in the regular region. In Fig. 5a 
we show the idempotency defect (linear entropy) for the initial conditions 
of the form (\ref{ic}) centered at the triangles marked on the Poincar\'e 
section of Fig. 1a, and in Fig. 5b the projection of the orbits for the specific 
cases of the initial conditions labeled $c_1$ and $c_2$ (see figure captions). In
these cases the linear entropy does not exhibit periodic oscillations, albeit
oscillations are still present. Although they are not directly   related to a periodic
orbit, one can explain the oscillations very much  along the lines followed in the
regular case. If we take a classical   trajectory associated with the center of the
wave packet and follow its   projection on the atomic phase space, we notice again
that the proximity  of this phase space border plays a decisive role on the
decoherence process. This is illustrated in Figs. 5c and 5d for the initial conditions
$c_1$ and $c_2$. For the case $c_2$, decoherence is retarded due to the fact that the
wave packet moves initially towards the center of the phase space, following the
classical trajectory shown in Fig. 5b. Note that as soon as the classical trajectory
approaches the border there is a corresponding increase in the entanglement rate
$\frac{d\delta}{dt}$ in Fig. 5d. Thus the linear entropy increases. In terms of the
representation independent quantity $\langle J_z \rangle (t)$, the same features
prevail. The remaining initial condition $c_3$ for which $\delta(t)$ has been evaluated
in Fig. 5a behaves in qualitatively analogous way, as for instance the trajectory
labeled $c_1$ shown in Fig. 5b, and will not be shown here. As to be expected, in the
chaotic region the relation between the motion of the quantum wave packet and the
classical trajectory associated to its center is rapidly overwhelmed by other effects
such as the rapid spreading of the quantum wave function. In fact, in Fig. 6 we show
the time evolution of the spin Husimi distribution associated with the initial
condition $c_2$ in Fig. 5. This figure illustrates the given argument. Moreover, we
see that for times when the plateau in $\delta(t)$ is reached, the spin
Husimi distribution is totally delocalized in the corresponding phase space. Finally we
note that this is not a characteristic of the chaotic initial condition;
the wave packets centered at regular initial conditions also become
spread all over the spin phase space by the time the linear entropy reaches
the plateau. 
 
\section{SUMMARY} 
 
  The present work has been devoted to a detailed analysis of the 
dynamics of the process of entanglement in the chaotic $N$-atom Jaynes-Cummings model. 
Previous work on the model \cite{nos,angelo99} pointed out general features of the
process, mainly focusing on differences between chaotic and regular regimes (unstable
and stable). The entanglement process of the model is however very rich and much more
can be learned from specific features of the linear entropy such as its oscillations
(recoherences in time, periodic or not). We have found an intimate connection between
these   recoherences and classical orbits. Specifically, we have pointed out the very
special role played by the morphology of the spin-projected classical  orbits. This 
is in accordance with the known fact that the decoherence  properties are dictated 
by the smaller subsystem \cite{smaller}. 

  Considering the present state of experimental proposals and realizations of 
atomic coherent states in the $N$-atom JCM, it seems that the integrable RW 
case ($G^{'}=0$) is more likely to be realizable in a practical implementation 
of the model considered in this work. Although not shown here, the RW case was 
checked for regular initial conditions on tori in phase space and showed similar 
recoherence phenomena, as described in the previous sections.
 
  Despite the fact that these findings are model dependent, we believe them to be 
typical of systems involving two degrees of freedom, one of them having a Hilbert 
space with dimension much greater than the dimension of the other, and whose classical 
limit is chaotic (soft chaos). 
 
\begin{center} 
{\bf Acknowledgements} 
\end{center} 
\par It is a pleasure to thank Prof. H. A. Weidenm\"uller for first calling our 
attention to the need to explain the oscillations in the linear entropy, and 
T. H. Seligman for helpful discussions and for reading the 
manuscript.  The authors acknowledge financial support to the Brazilian agencies 
Funda\c{c}\~ao de Amparo \`a Pesquisa do Estado de S\~ao Paulo (FAPESP) and 
Conselho Nacional de Desenvolvimento Cient\'{\i}fico e Tecnol\'ogico (CNPq).

\newpage 
\newcounter{figura} 
\noindent 
{\bf Figure Captions} 
\hspace*{\parindent} \\ 
\begin{list} 
{Figure \arabic{figura}:}{\usecounter{figura} 
\setlength{\rightmargin}{\leftmargin}} 
\item {\bf(a)} Poincar\'e section given by $q_f=0.0$ and $p_f> 0.0$ for the spin 
degree of freedom with $J=9/2$ in the nonintegrable ($G = 0.5$ and $G' = 0.2$) and
resonant ($\varepsilon = w_0=1$) case --- these values will be the same in all
subsequent figures except when otherwise noted. Here we show the section for the
energy $E=8.5$. The marks represent the various choices for the center of the coherent
states: circles for regular initial conditions (IC) and triangles for chaotic ones.  
{\bf(b)} Spin projection of the two shortest stable periodic orbits indicated by
circles in (a) (the border line is also shown). Initial conditions at the surface of
section are as follows:
$(q_a=0.0,p_a=2.261,q_f=0.0,p_f=3.423276)$ for the first (elliptic) orbit with period
$\tau = 4.89$; $(q_a=0.0,p_a=-3.577,q_f=0.0,p_f=5.221656)$ for the second orbit with
period $\tau = 7.45$.  
 
\item {\bf(a)}Poincar\'e section for the energy $E=35$. The circle represents the 
center of the coherent state for one regular initial condition. {\bf(b)} Spin
projection of the shortest stable periodic orbit, indicated by the circle in (a).
Initial condition at the surface of section is as follows:
$(q_a=0.0,p_a=1.4175,q_f=0.0,p_f=7.888904)$ with period $\tau = 5.82$.  
 
\item Linear entropy $\delta(t)$ (continuous lines), its derivative 
$\frac{d \delta(t)}{dt}$ (dashed lines), and the normalized expectation value $\langle
J_z \rangle (t)/J$ (dot-dashed lines) as functions of time. {\bf(a)} for the first
regular initial condition of Fig. 1b. Note the correspondence between the maxima and
minima of $\frac{d \delta(t)}{dt}$ and $\langle J_z \rangle (t)/J$. {\bf(b)} for the
second regular initial condition of Fig. 1b. In this case, only the minima of $\frac{d
\delta(t)}{dt}$ and $\langle J_z \rangle (t)/J$ have a clear correspondence due to the
peculiar form of the orbit (see second figure of Fig. 1b). {\bf(c)} Regular initial
condition as marked in Fig. 2a and its spin-projected orbit shown in Fig. 2b; during
the interval of time corresponding to the rise of $\delta(t)$ there is an
approximate   correspondence between the maxima and minima of $\frac{d \delta(t)}{dt}$
and those of $\langle J_z \rangle (t)/J$. 
  
\item For the energy $E=8.5$ in the integrable ($G = 0.5$ and $G' = 0.0$)
and resonant ($\varepsilon = w_0=1$) case, the plot of linear entropy
$\delta(t)$ for initial  condition on a periodic orbit which is {\bf
circular} (see inset)
$(q_a=0.0,p_a=2.47675,q_f=0.0,p_f=3.563642)$. Notice that the curve shows practically
no oscillations as compared to Fig. 3a. 
 
\item {\bf(a)} Linear entropy $\delta(t)$ for the {\bf chaotic initial conditions} 
corresponding to the triangles shown in Fig. 1a: $c_1$
$(q_a=-4.0,p_a=0.0,q_f=0.0,p_f=3.162278)$ for the continuous line; $c_2$
$(q_a=1.57,p_a=-2.0,q_f=0.0,p_f=5.680465)$ for the dashed line; and $c_3$
$(q_a=3.0,p_a=2.0,q_f=0.0,p_f=2.942413)$ for the dot-dashed line. {\bf(b)} Spin
projections of the orbits {\bf $c_1$} and {\bf $c_2$}. Various times are indicated
along the orbits showing that the initial conditions are such that the trajectory {\bf
$c_2$} in spin phase space is launched inward as opposite to the case {\bf $c_1$} for
instance. {\bf(c)} $\delta(t)$ (continuous line), $\frac{d \delta(t)}{dt}$ (dashed
line), and $\langle J_z \rangle (t)/J$ (dot-dashed line) for {\bf $c_1$}. {\bf(d)}
$\delta(t)$ (continuous line), $\frac{d \delta(t)}{dt}$ (dashed line), and $\langle
J_z \rangle (t)/J$ (dot-dashed line) for {\bf $c_2$}. During the interval of time
corresponding to the rise of $\delta(t)$ there is an approximate correspondence 
between the maxima and minima of $\frac{d \delta(t)}{dt}$ and $\langle J_z \rangle
(t)/J$ but for shorter time than it happens for the regular cases shown in Figs. 3a
and 3b. 
 
\item Spin Husimi distribution of the quantum coherent state initially centered at 
condition {\bf $c_2$} in Fig. 5. Snapshots are taken at time values $t = 0$ (upper
left), $t = 1$ (upper right), $t = 4$ (down left), and $t = 25$. 
 
\end{list} 

\newpage  
\begin{thebibliography}{99}
 
\bibitem{Einstein} Letter from Albert Einstein to Max Born in 1954, cited 
by E. Joos, in {\it New Techniques and Ideas in Quantum Measurement Theory},  
edited by D. M. Greenberger (New York Academy of Science, New York, 1986).
 
\bibitem{decoherence} H. D. Zeh, Found. Phys. {\bf 1}, 69 (1970); H. Dekker, Phys. 
Rev. A {\bf 16}, 2126 (1977); W. H. Zurek, Phys. Rev. D {\bf 24}, 1516 (1981); {\bf
26}, 1862 (1982); W. G. Unruh and W. H. Zurek, {\it ibid.} {\bf 40}, 1071 (1989); W. H.
Zurek, Phys. Today {\bf 44}, 36 (1991); B. L. Hu, J. P. Paz, and Y. Zhang, Phys. Rev.
D {\bf 45}, 2843 (1992); W. H. Zurek, S. Habib, and J. P. Paz, Phys. Rev. Lett. {\bf
70}, 1187 (1993); D. Giulini, E. Joos, C. Kiefer, J. Kupsch, I. O. Stamatescu, and H.
D. Zeh, {\it Decoherence and the Appearance of a Classical World in Quantum Theory}
(Springer-Verlag, Berlin, 1996). 

\bibitem{Brune} M. Brune, E. Hagley, J. Dreyer, X. Ma\^{\i}tre, A. Maali,  
C. Wunderlich, J. M. Raimond, and S. Haroche, Phys. Rev. Lett. {\bf 77}, 4887 (1996); 
C. Monroe, D. M. Meekhof, B. E. King, and D. J. Wineland, Science {\bf 272}, 1131
(1996).  

\bibitem{zurek95} W. H. Zurek and J. P. Paz, Physica (Amsterdam) {\bf 83D},  
300 (1995); W. H. Zurek, S. Habib, and J. P. Paz, Phys. Rev. Lett. {\bf 70},  
1187 (1993).  

\bibitem{Shiokawa} A. Tameshtit and J. E. Sipe, Phys. Rev. A {\bf 47}, 1697 (1993); 
K. Shiokawa and B. L. Hu, Phys. Rev. E {\bf 52}, 2497 (1995).
 
\bibitem{nos} K. Furuya, M. C. Nemes, and G. Q. Pellegrino, Phys. Rev. Lett. 
{\bf 80}, 5524 (1998).

\bibitem{NJCM} M. Tavis and F. W. Cummings, Phys. Rev. {\bf 170}, 379 (1968); 
see also R. H. Dicke, {\it ibid.} {\bf 93}, 99 (1954). 

\bibitem{ATOMCAT} G. S. Agarwal, R. R. Puri and R. P. Singh, Phys. Rev. A {\bf 56}, 
2249 (1997); C. C. Gerry and R. Grobe, {\it ibid.} {\bf 56}, 2390 (1997);
{\bf 57}, 2247 (1998); M. G. Benedict and A. Czirj\'ak, {\it ibid.} {\bf 60},
4034 (1999). 

\bibitem{Recamier00} J. Recamier, O. Casta\~nos, R. J\'auregui, and A. Frank, 
Phys. Rev. A {\bf 61}, 063808 (2000).

\bibitem{Rauschenbeutel00} A. Rauschenbeutel, G. Nogues, S. Osnaghi, P. Bertet, 
M. Brune, J.-M. Raimond, and S. Haroche, Science {\bf 288}, 2024 (2000).

\bibitem{Sorensen00} K. M{\o}lmer and A. S{\o}rensen, Phys. Rev. Lett. {\bf 82}, 
1835 (1999); A. S{\o}rensen and K. M{\o}lmer, Phys. Rev. A {\bf 62}, 022311 (2000).

\bibitem{Turchette98} Q. A. Turchette, C. S. Wood, B. E. King, C. J. Myatt, 
D. Leibfried, W. M. Itano, C. Monroe, and D. J. Wineland , Phys. Rev. Lett. 
{\bf 81}, 3631 (1998); C. A. Sackett, D. Kielpinski, B. E. King, C. Langer, 
V. Meyer, C. J. Myatt, M. Rowe, Q. A. Turchette, W. M. Itano, D. J. Wineland, 
and C. Monroe, Nature (London) {\bf 404}, 256 (2000).

\bibitem{Roos99} Ch. Roos, Th. Zeiger, H. Rohde, H. C. N\"agerl, J. Eschner, 
D. Leibfried, F. Schmidt-Kaler, and R. Blatt, Phys. Rev. Lett. {\bf 83}, 
4713 (1999).

\bibitem{Cirac96} J. I. Cirac, A. S. Parkins, R. Blatt, and P. Zoller, 
Adv. Atom. Mol. Phys. {\bf 37}, 237 (1996).

\bibitem{Milonni98} P. W. Milonni, J. R. Ackerhalt and H. W. Galbraith, Phys. Rev. 
Lett. {\bf 50}, 966 (1983); P. W. Milonni, M. L. Shih and J. R. Ackerhalt, {\em
Chaos in Laser-Matter Interactions}, World Scientific Lecture Notes in Physics
Vol.6 (World Scientific, Singapore, 1987); R. Graham and M. H\"ohnerbach, 
Z. Phys. B {\bf 57}, 233 (1984); idem in {\em Quantum Measurement and Chaos}, 
edited by E. R. Pike and S. Sarkar, NATO Advanced Study Institute, Series B, 
Vol. 161, (Plenum, New York, 1987), p. 147.

\bibitem{klauder85} J. R. Klauder and B.-S. Skagerstam, {\it Coherent States: 
Applications in Physics and Mathematical Physics} (World Scientific, Singapore, 1985).
 
\bibitem{aguiar92} M. A. M. de Aguiar, K. Furuya, C. H. Lewenkopf, and M. C. Nemes, 
Ann. Phys. {\bf 216}, 291 (1992). 

\bibitem{lewenkopf91} C. H. Lewenkopf, M. C. Nemes, V. Marvulle, M. P. Pato, and 
W. F. Wreszinski, Phys. Lett. A {\bf 155}, 113 (1991). 

\bibitem{aguiar91} M. A. M de Aguiar, K. Furuya, C. H. Lewenkopf, and M. C. Nemes, 
Europhys. Lett. {\bf 15}, 125 (1991); K. Furuya, M. A. M de Aguiar, C. H. Lewenkopf,
and M. C. Nemes, Ann. Phys. {\bf 216}, 313 (1992). 

\bibitem{note} This connection is easily seen by noting that 
$\brahket{wv}{J_z}{wv}~=~\frac{1}{2}\left(p^2_a + q^2_a - 2J \right)$. 

\bibitem{angelo99} R. M. Angelo, K. Furuya, M. C. Nemes, and G. Q. Pellegrino, Phys. 
Rev. E {\bf 60}, 5407 (1999). 

\bibitem{smaller} E. Schr\"odinger, Proc. Phil. Soc. {\bf 31}, 555 (1935); {\bf 32},
446 (1936); M. C. Nemes and A. F. R. de Toledo Piza, Physica A  {\bf 137}, 367 (1986);
A. Ekert and P. L. Knight, Am. J. Phys. {\bf 63}, 415 (1995).  

\end{thebibliography}
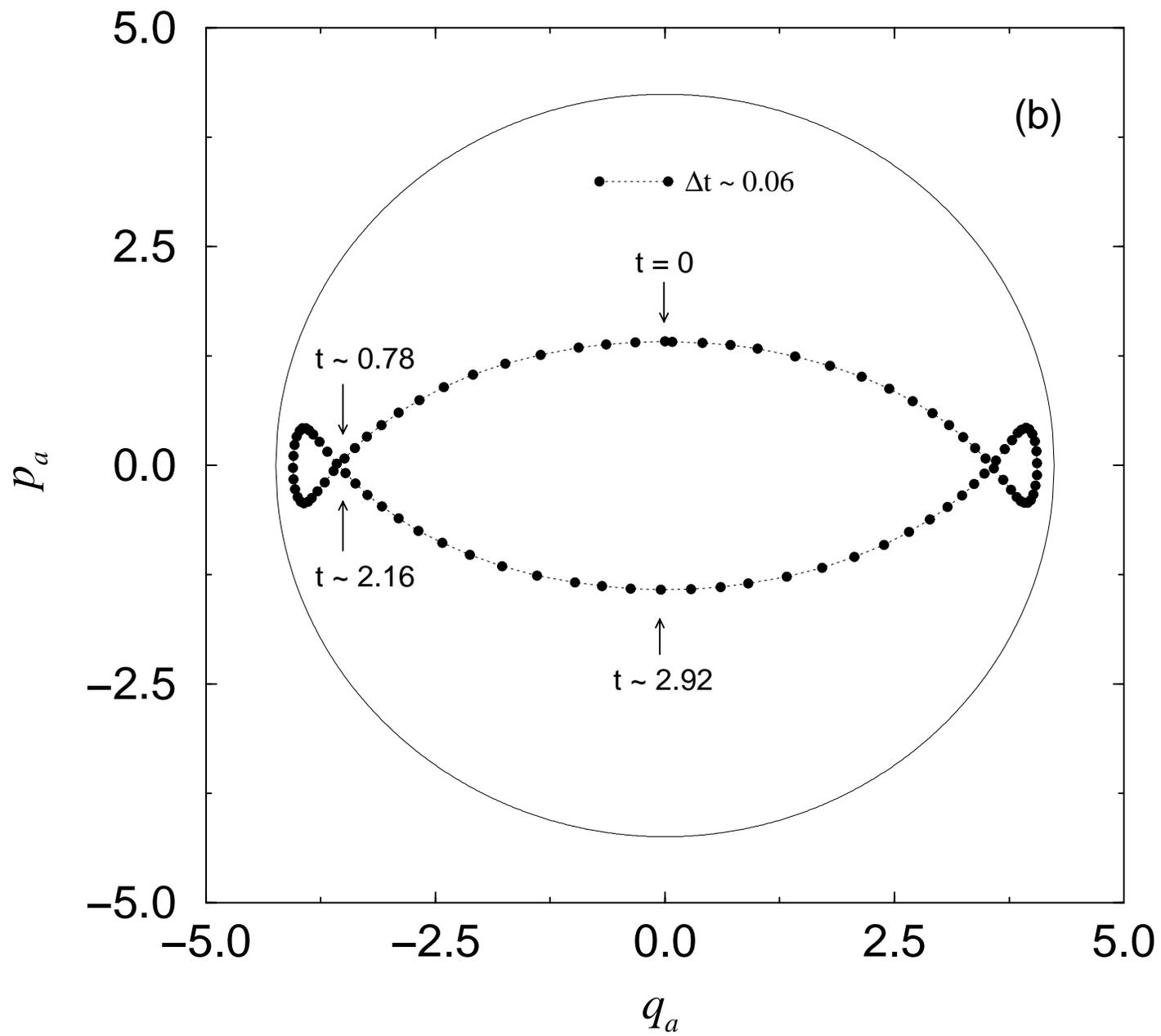
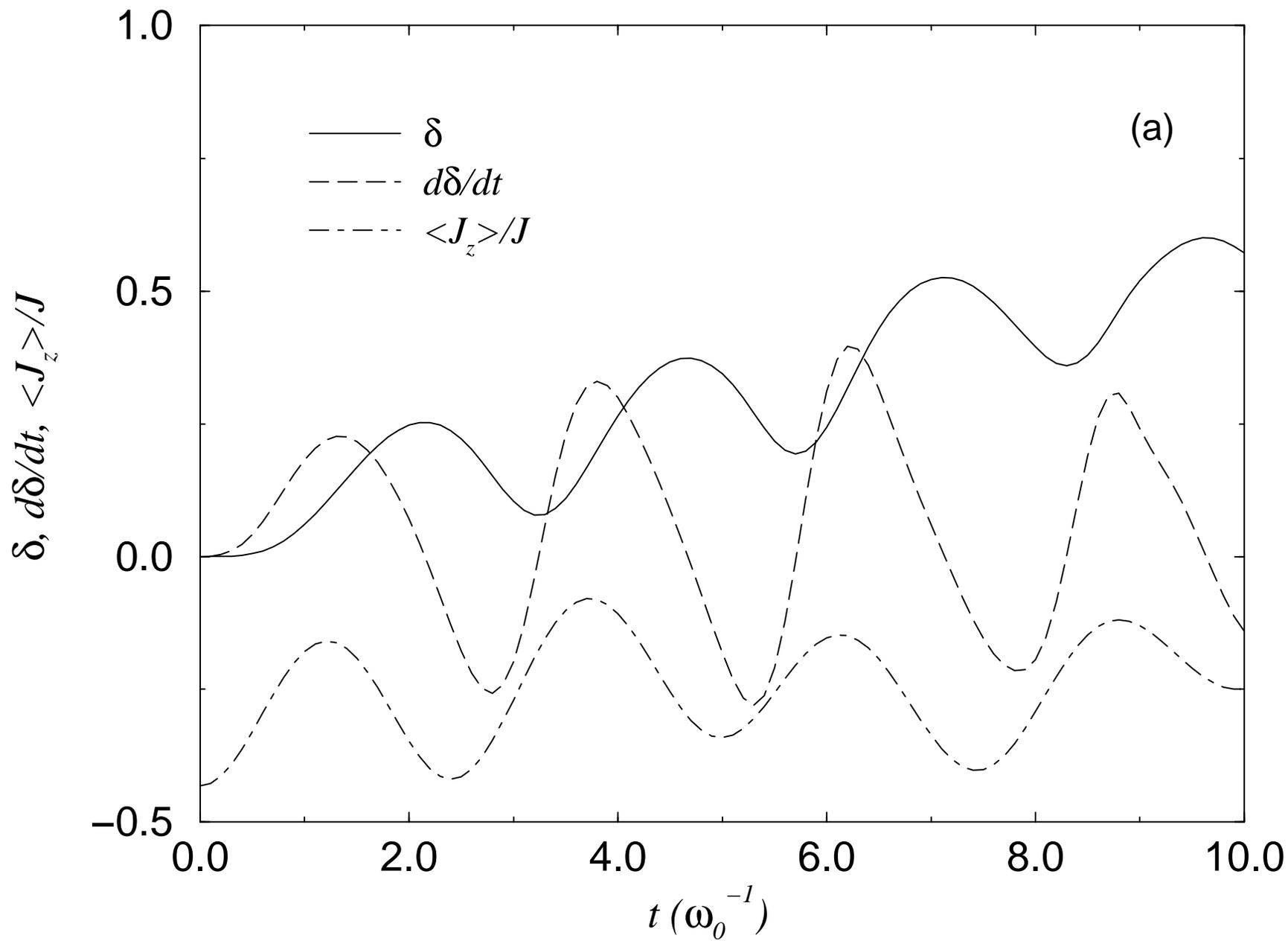
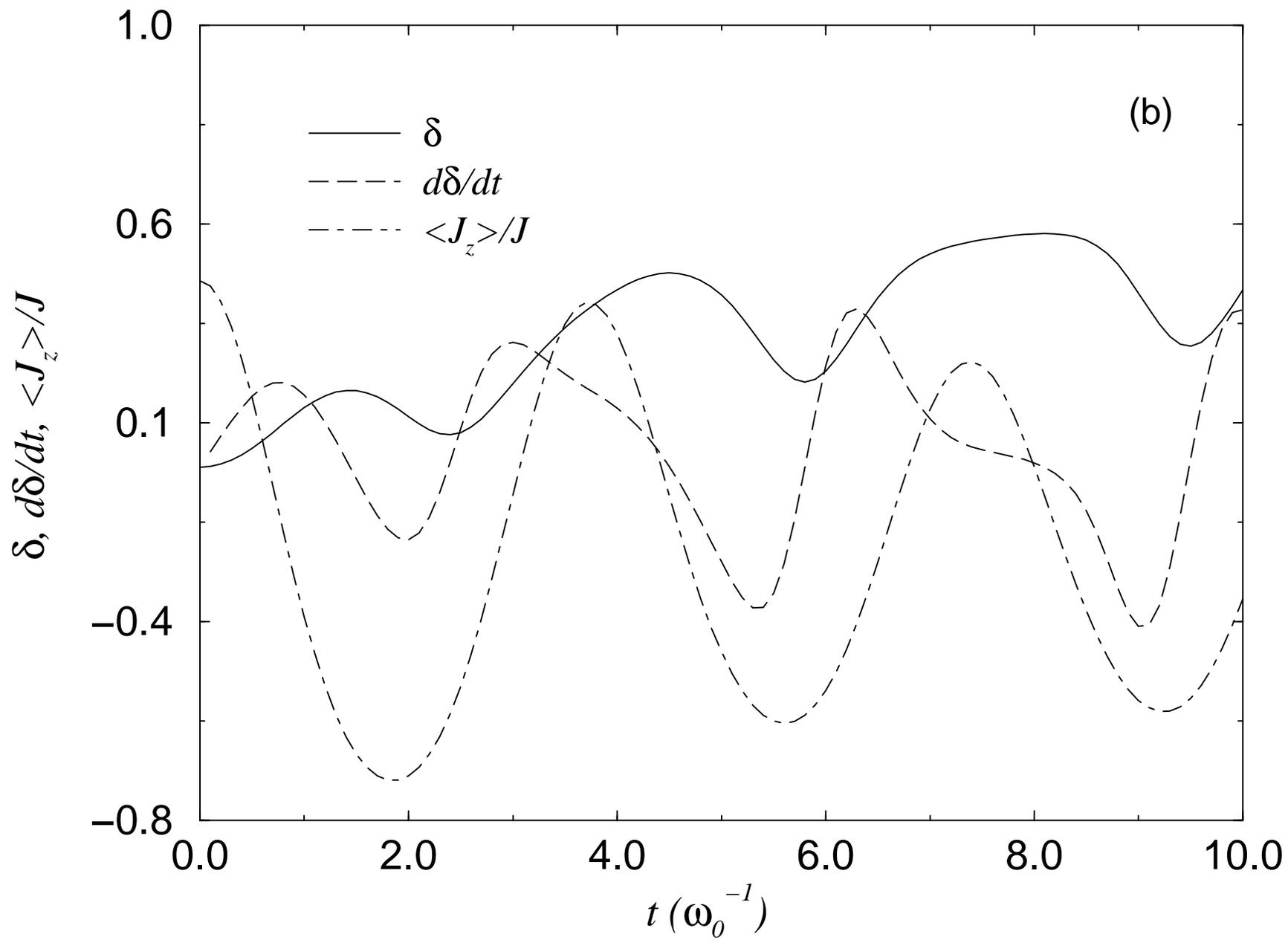
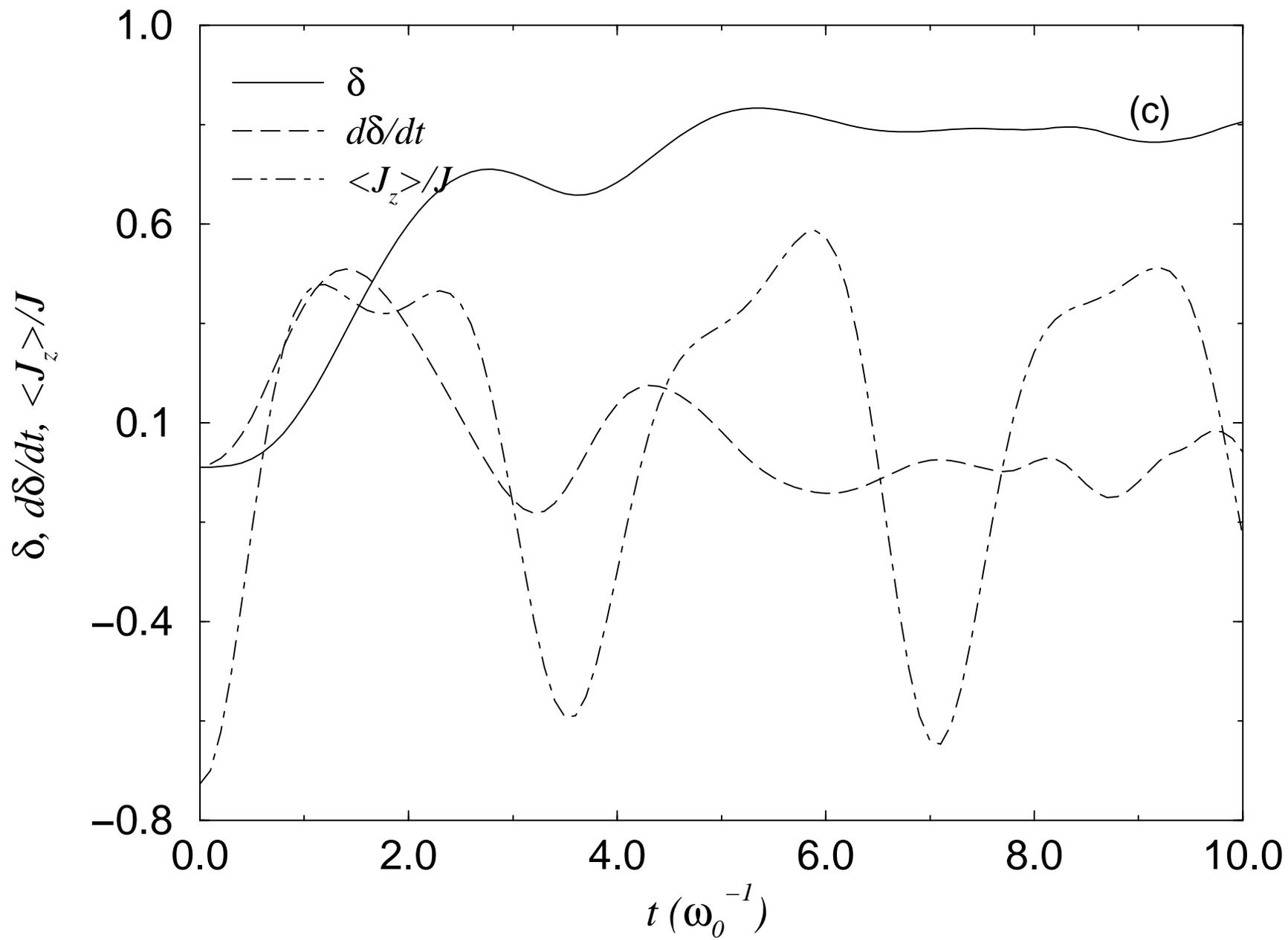
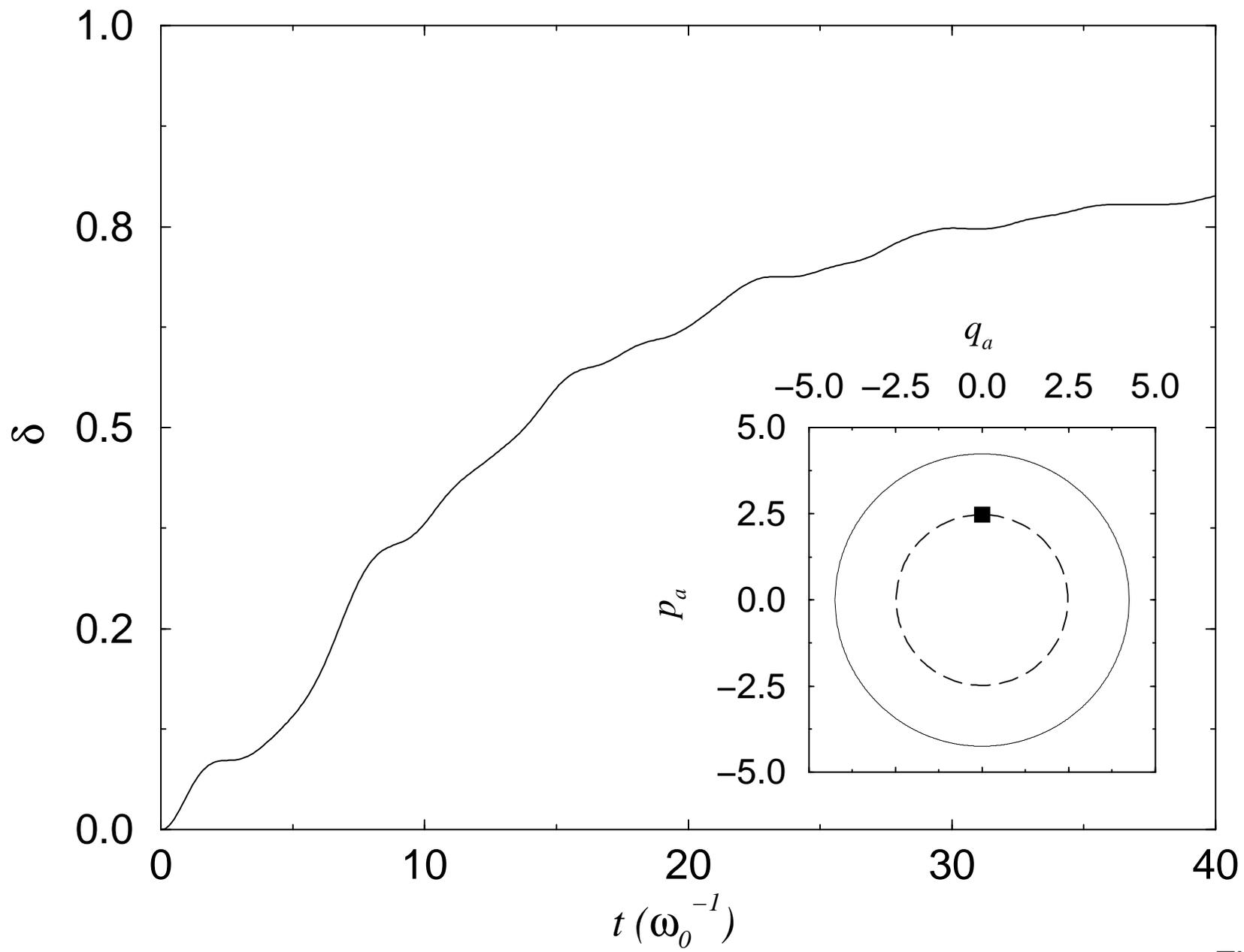
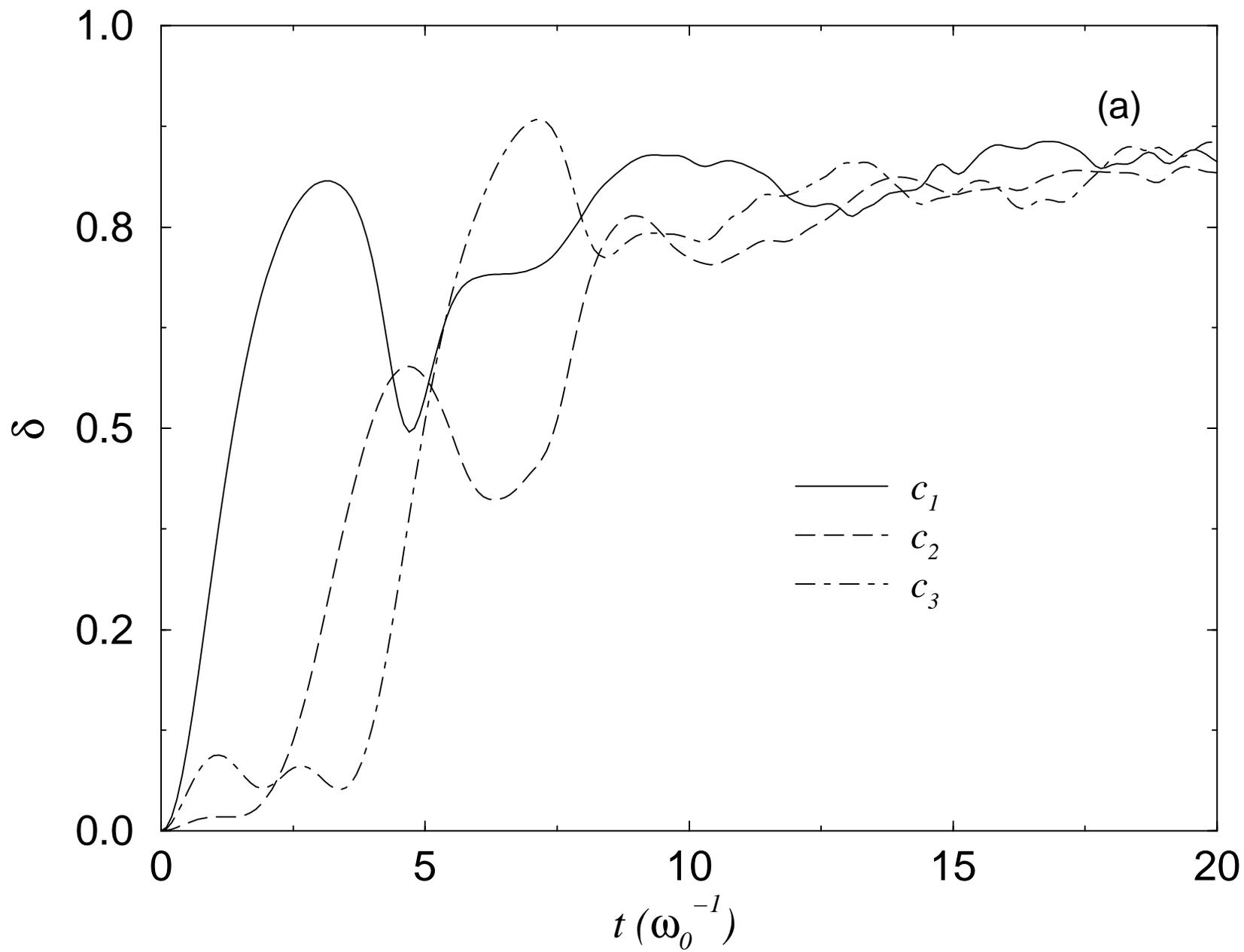
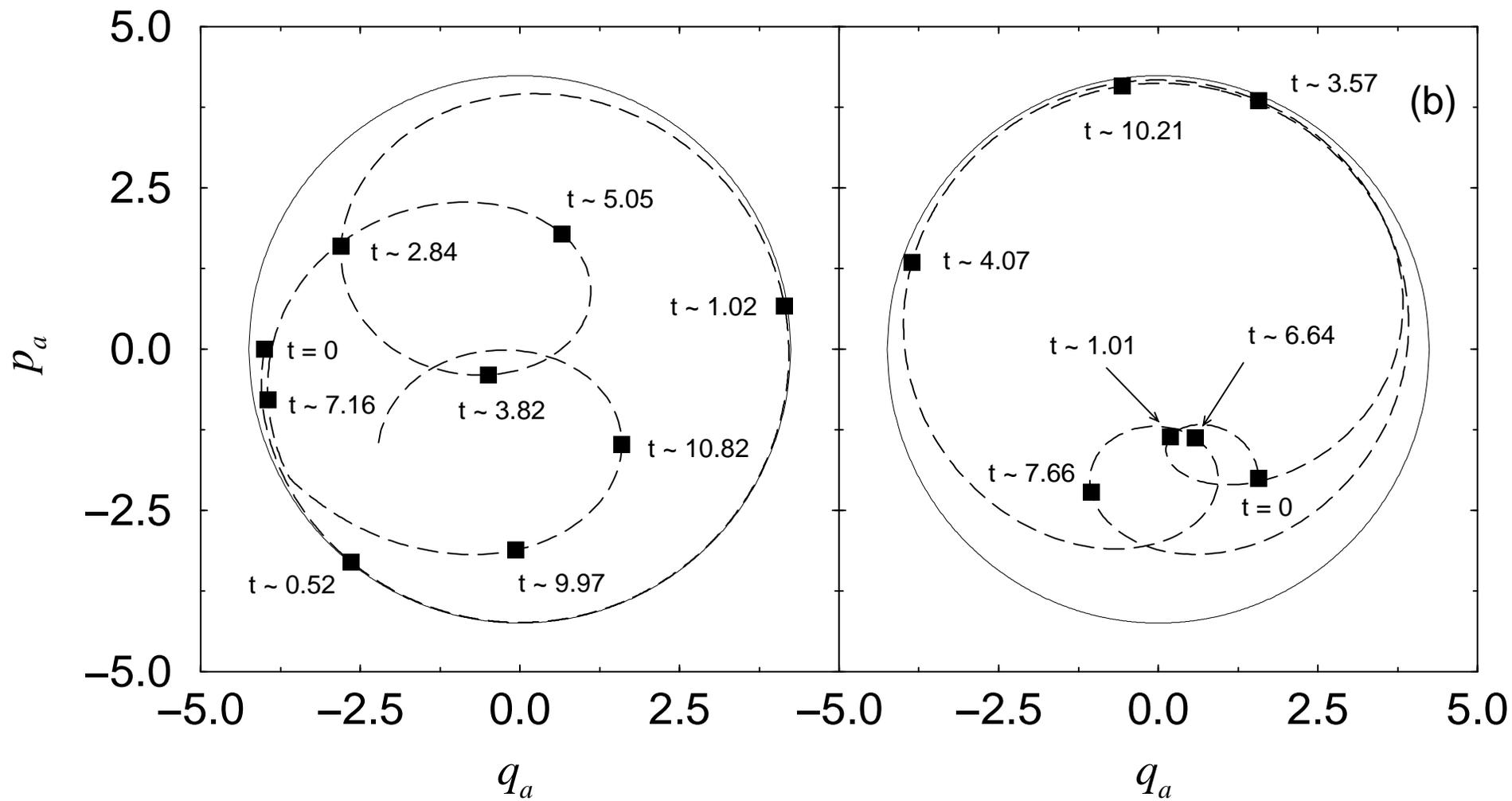
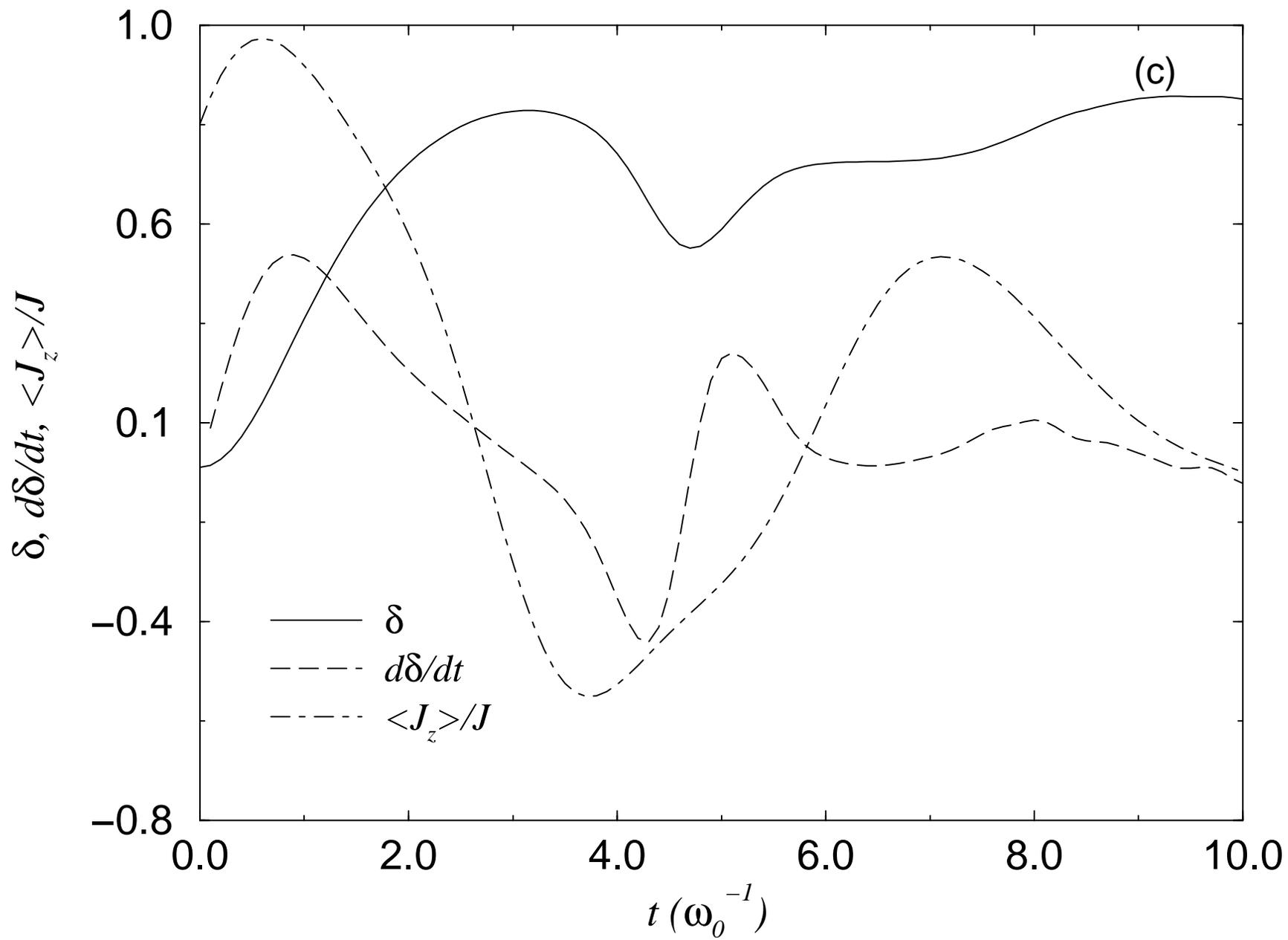
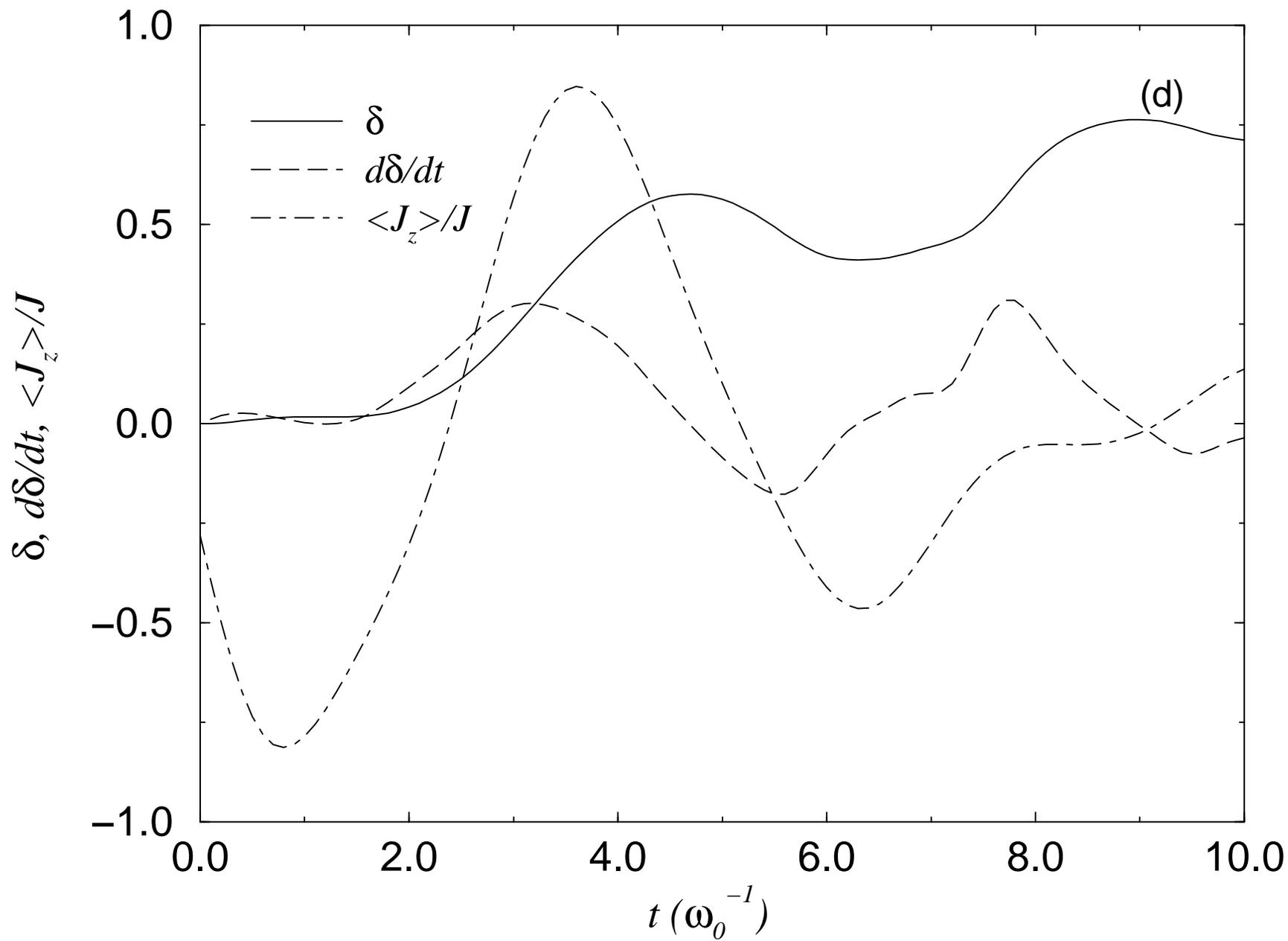
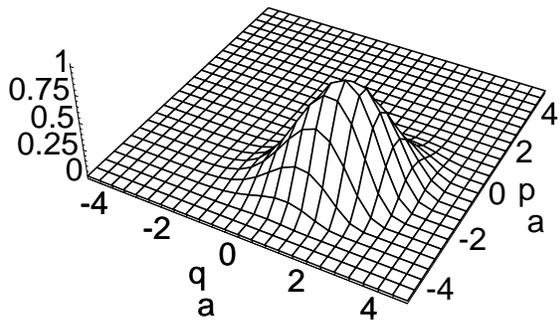
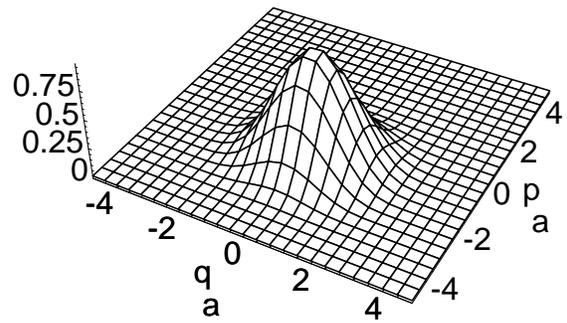
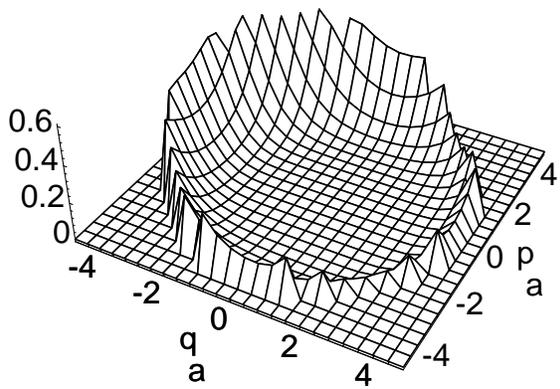
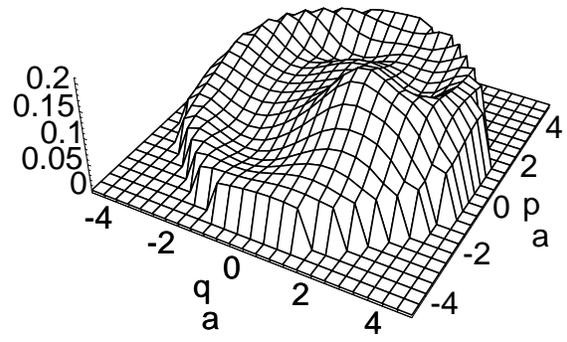
\end{document}